\documentclass{article}

\usepackage{amsmath,amsfonts,amssymb,amsthm}
\usepackage{setspace,graphicx}
\usepackage{natbib}
\usepackage{bm}
\usepackage{ifthen}
\usepackage{verbatim}
\usepackage{color}
\usepackage{dsfont}
\usepackage{tikz}
\usepackage{setspace,graphicx}
\usepackage{sgame}
\usepackage{bm}
\usepackage{multirow}
\usepackage{ifthen}
\usepackage{verbatim}
\usepackage{xcolor}
\usetikzlibrary{arrows.meta}

\theoremstyle{plain}
\theoremstyle{definition}

  \newtheorem{corollary}{Corollary}
  \newtheorem{definition}{Definition}
  
  \newtheorem{example}{Example}
  \newtheorem{lemma}{Lemma}
  
  \newtheorem{proposition}{Proposition}

  \theoremstyle{remark}

\def\EE{\mathbb{E}}
\def\RR{\mathbb{R}}
\DeclareMathOperator*{\argmax}{arg\,max}
\def\XX{\mathcal{X}}

\begin{document}


\title{Marginal stochastic choice}

\author{Yaron Azrieli and John Rehbeck\thanks{Department of Economics, The Ohio State University, 1945 North High street, Columbus, OH 43210.
azrieli.2@osu.edu and rehbeck.7@osu.edu}}

\maketitle

\begin{abstract}
Models of stochastic choice typically use conditional choice probabilities given menus as the primitive for analysis, but in the field these are often hard to observe. Moreover, studying preferences over menus is not possible with this data. We assume that an analyst can observe marginal frequencies of choice and availability, but not conditional choice frequencies, and study the testable implications of some prominent models of stochastic choice for this dataset. We also analyze whether parameters of these models can be identified. Finally, we characterize the marginal distributions that can arise under two-stage models in the spirit of \citet{gul2001temptation} and of \citet{kreps1979representation} where agents select the menu before choosing an alternative.




\end{abstract}

\newpage
\section{Introduction}\label{sec-introduction}

The vast majority of papers on stochastic choice assume that the data available to the observer contains choice frequencies conditional on a wide collection of choice sets (menus). Indeed, the early works of \cite{luce1959individual}, \cite{Block1960},  \cite{falmagne1978representation}, \cite{barbera1986falmagne}, and \cite{mcfadden1990stochastic} demonstrated the potential of this framework to deliver elegant and intuitive characterizations of choice rules, and subsequent works followed this tradition.\footnote{More recent papers that use conditional choice probabilities to characterize various decision models include \cite{gul2006random}, \cite{manzini2014stochastic}, \cite{fudenberg2015dynamic}, \cite{fudenberg2015stochastic}, \cite{brady2016menu}, \cite{aguiar2017random}, \cite{apesteguia2017single}, \cite{kitamura2018nonparametric}, \cite{frick2019dynamic},  \cite{cattaneo2020random}, \cite{cattaneo2021attention}, and \cite{kovach2022behavioral}. See \cite{strzalecki2022sct} for a comprehensive survey of the stochastic choice literature.} In practice, however, focusing on conditional choice frequencies has two important drawbacks. First, in the field it is often hard to observe the menu that an individual faced, and therefore data on conditional probabilities may be hard to find. Second, in many cases decision makers choose the menu before choosing an alternative, but studying preferences over menus requires data on the frequency with which menus are selected.

To address these problems this paper introduces and studies \emph{marginal stochastic choice datasets}. Such datasets consist of a pair of distributions, one over menus and one over alternatives. Throughout the paper we denote the distribution over menus by $\mu$ and the distribution over alternatives by $\lambda$. Here, $\mu(A)$ is the share of choices made in the $A$th menu, while $\lambda(a)$ is the aggregate frequency that the $a$th alternative is chosen. We revisit some of the prominent models in the stochastic choice literature that were characterized using conditional choice frequencies, and study their properties with this new dataset. In addition, we study the testable implications of the models of \citet{gul2001temptation} and of \citet{kreps1979representation} where agents have preferences over both menus and alternatives.

Marginal stochastic choice datasets naturally arise when a researcher only has access to aggregate-level data. For example, research on grocery store purchases often uses market-level (e.g.\ city-level) data, but consumers who visit different stores may have different options to choose from. Here, observing $\lambda$ means that the data contains the aggregate market share of each product (but not store-level sales). Observing $\mu$ means that the set of stocked products and share of consumers that shops at each grocery store can be estimated.\footnote{\cite{rhee2002inter} shows that many consumers shop at only one or two grocery stores. \cite{hausman1994competitive}, \cite{nevo2001measuring}, and \cite{hausman2002competitive} estimate demand in various grocery categories assuming that all consumers face the same set of options. \cite{tenn2008biases} shows that not accounting for retail distribution may generate biases in product price elasticises.}

We say that the marginal stochastic choice dataset $(\mu,\lambda)$ is rationalizable with a given model of stochastic choice when the marginal distribution of choices ($\lambda$) can be obtained from the marginal distribution of menus ($\mu$) and from conditional choice probabilities consistent with the model using the law of total probability. As a benchmark, we start with the case where no restrictions are imposed on the way agents choose given the menu they face. Even here not every $(\mu,\lambda)$ is rationalizable, because alternatives that are rarely available cannot be frequently chosen. More specifically, $(\mu,\lambda)$ is rationalizable if and only if
\begin{equation}\label{eqn-intro}
    \sum_{ a \in A} \lambda(a) \ge \sum_{B \subseteq A} \mu(B) \tag{$*$}
\end{equation}
for every menu $A$. The inequality (\ref{eqn-intro}) is a variant of the condition in the classic `marriage lemma' \citep{hall1934representation} and simply means that the frequency with which alternatives in $A$ are chosen (the left-hand side) must be at least as large as the frequency with which only alternatives in $A$ are available (the right-hand side).

We proceed by imposing more structure on the decision process. We find that assuming choices are made according to the random utility model \citep{Block1960, falmagne1978representation} places no additional restrictions on marginal stochastic choice data beyond the collection of inequalities (\ref{eqn-intro}). This is in contrast to the case in which conditional frequencies are observable where the model does have testable implications. Further, the distribution of preferences typically cannot be identified.\footnote{It is well-known that the distribution of preferences in the random utility model is not identified even when conditional choice frequencies are observable \citep{falmagne1978representation, fishburn1998stochastic}. See Section \ref{sec-RUM} for a discussion of the differences in identification between the two datasets.} However, we show how one can obtain information on features of the distribution of preferences from marginal stochastic choice data.

Next, we find that even the strong assumption of having choices generated by the Luce model \citep{luce1959individual} essentially places no additional restrictions on marginal stochastic choice data. However, unlike with random utility, under Luce the parameter is identified as long as the support of $\mu$ satisfies a mild condition. Thus, under the popular Luce model, if the researcher has information about the distribution of menu availability $\mu$, then it is enough to observe aggregate choices in order to pin down conditional choices.

Lastly, we find that the independent random consideration model of \cite{manzini2014stochastic} places non-trivial restrictions on the ratios of aggregate choice probabilities and menu probabilities. Moreover, we show that if the underlying preference order is known then the consideration probability for each alternative can be recovered from the marginals. If the preference is unknown, then identification fails in general, but one can identify both the preference and the consideration probabilities whenever the outside option is chosen with sufficiently low probability.

In the models considered so far we viewed the distribution of menu availability $\mu$ as exogenously given.\footnote{This distribution is clearly an important part of the dataset because it affects what distributions of choice $\lambda$ can emerge under a given model of behavior, but up to this point $\mu$ was not influenced by or influencing how agents choose.} But in practice agents often choose the menu themselves before choosing an alternative from the menu, e.g., consumers may choose a grocery store partly based on the products available at that store. In the last part of the paper we consider two-stage models consisting of a first-period menu choice and a second-period alternative choice. Thus, both $\mu$ and $\lambda$ are now endogenous. Furthermore, the choice of a menu not only determines what alternatives are available in the second period, but is also informative about the preferences over alternatives, and therefore about the alternative that the agent eventually chooses. We demonstrate the potential of this approach by characterizing the datasets $(\mu,\lambda)$ that can be rationalized by the temptation and self-control model of \cite{gul2001temptation}, as well as by the model of \citet{kreps1979representation} in which agents have preference for flexibility. It turns out that these models place intuitive restrictions on marginal stochastic choice data. For example, in the temptation and self-control model the key is that an agent would never choose a menu $A$ and then an alternative $a$ from that menu if there is a feasible sub-menu of $A$ that also contains $a$. This property translates into a restriction on the support of $\mu$ as well as a strengthening of the inequalities in (\ref{eqn-intro}).

From a technical point of view our analysis heavily relies on classic results and ideas from the theory of transferable utility cooperative games. The connection is that instead of describing the distribution of availability by its probability mass function $\mu$, we can describe it by its cumulative distribution function $v_\mu(A)=\sum_{B\subseteq A} \mu(B)$. The inequalities (\ref{eqn-intro}) can then be understood as requiring that $\lambda$ is in the core of the game $v_\mu$. We leverage known results about the core to derive several of our characterizations. While some connections between cooperative games and stochastic choice have already been pointed out in the literature,\footnote{See for example \cite{gilboa1992game} and the references therein.} we believe that the connections we uncover may be useful in future work on stochastic choice.

\subsection{Literature review}\label{subsec-lit_rev} 

The recent work of \cite{dardanoni2020inferring} is motivated by similar considerations to the current paper, namely, that theoretical models of stochastic choice should be based on datasets more likely to be available in the field. In their model agents are heterogeneous in their cognitive ability which determines the number of alternatives they can consider. The researcher observes the aggregate choice distribution ($\lambda$ of the current paper) and the question is whether the distribution of cognitive types can be identified. Thus, roughly speaking, the main difference from our work is that we assume that the distribution of menus is observable and study conditional choices, while they study whether the menu distribution can be inferred from aggregate choices.\footnote{The literature has interpreted the menu from which an agent chooses either as the set of feasible alternatives or as the `consideration set' of alternatives the agent pays attention to. In our case the first interpretation is more appropriate since we assume that the distribution of menus is observable.} 

There is a long history of empirical and econometrics papers trying to deal with the problem of unobservable choice sets. This problem is discussed in length in \cite{manski1977structure}, while some examples of papers on related issues include \cite{swait1986analysis}, \cite{horowitz1991modeling}, \cite{ben1995discrete}, \cite{tenn2008biases}, \cite{tenn2009demand}, and \cite{abaluck2021consumers}. Some of these papers allow for menus to be at least partly chosen by the decision makers, similar to the setup we consider in Section \ref{sec-2-period}. Recently, \cite{barseghyan2021heterogeneous} study a random utility model where the distribution of availability is not observed but there is a known lower bound on the size of menus. Their characterization of the sharp identification region in Theorem 3.1 has some similarities to our Proposition \ref{prop-basic}. 

This paper also contributes to the emerging literature taking insights from stochastic choice to new settings and datasets. \cite{chambers2021correlated} study correlated preferences within the random utility framework, where the dataset assumed to contain the frequencies of tuples of choices made by a group of agents conditional on the menu that was available to each one of them. \cite{manzini2019sequential} studies whether or not an individual chooses to approve an option from a list. This stochastic dataset is novel since the sum of approving different options can be more than one, and since choice probabilities vary with the order in which options are presented. Another novel dataset is from \cite{cheung2021decision}, where decision makers obtain a recommendation before choosing an alternative. This differs from the standard conditional choice probabilities since any alternative can be chosen regardless of what menu was recommended.

\section{Preliminaries}\label{sec-preliminaries}

For any finite set $Y$ we denote by $\Delta(Y)$ the set of probability distributions on $Y$. If $y\in Y$ and $p\in \Delta(Y)$ then we write $p(y)$ instead of $p(\{y\})$. We sometimes identify $\Delta(Y)$ with the standard simplex in $\RR^Y$, i.e.\ with the set of $p\in \RR^Y$ such that $p(y)\ge 0$ for every $y$ and $\sum_y p(y)=1$. We use $Int(D)$ to denote the relative interior of a set $D\subseteq \RR^Y$, where $\RR^Y$ is endowed with its standard topology. In particular, $Int(\Delta(Y))$ is the set of $p\in \Delta(Y)$ such that $p(y)>0$ for all $y$.

Throughout the paper we denote by $X$ the finite set of alternatives and by $\XX=2^X\setminus{\emptyset}$ the collection of all non-empty subsets of $X$. Alternatives in $X$ are typically denoted by $a,b,\ldots$, while elements of $\XX$ are called menus and are typically denoted by $A,B,\ldots$. The cardinality of $X$ is $|X|=n$. If $A\in \XX$ then $A^c=X\setminus A$ is the complement of $A$.

In this paper, we study what can be learned from the \emph{marginal stochastic choice dataset} $(\mu,\lambda) \in \Delta(\XX) \times \Delta(X)$, where $\mu$ is the frequency with which menus are available and $\lambda$ is the aggregate frequency of choices of alternatives. We sometimes refer to $\lambda$ as the marginal distribution over alternatives and $\mu$ as the marginal distribution over menus. We highlight that these are not choice frequencies conditional on a given menu as is common in standard models of stochastic choice.


Much of our analysis relies on classic results from the theory of cooperative games, so we now briefly present the essential definitions and results from that theory used in later sections. More details can be found in \cite{grabisch2016set}. At a high-level, we map the observed marginal menu distribution $\mu$ to a cooperative game and then use insights from cooperative game theory to characterize the model. 

A \emph{cooperative game} is a set function $v:2^X\to \RR$ satisfying $v(\emptyset)=0$. Throughout the paper we assume that any cooperative game is normalized so that $v(X)=1$. The core of a game $v$, denoted $Core(v)$, is the set 
$$Core(v)=\left\{p\in \RR^X :~ \sum_{a\in X} p(a)=1 \textit{ and } \sum_{a\in A}p(a)\ge v(A) ~~ \forall A\in \XX \right\}.$$

A game $v$ is \emph{convex} (or super-modular) when $v(A\cup B) + v(A\cap B) \ge v(A)+v(B)$
for all menus $A,B\in \XX$. A game is \emph{strictly convex} when the above inequality is strict whenever neither $A$ contains $B$ nor $B$ contains $A$. Given any real vector $z=(z(B))_{B\in \XX}$ with $\sum_{B \in \XX} z(B)=1$ we define the game $v_z$ by $v_z(A)=\sum_{B\subseteq A} z(B)$. Conversely, for any game $v$ there is a unique vector $z_v = (z_v(B))_{B\in \XX}$ such that $v(A)=\sum_{B\subseteq A} z_v(B)$ for every $A\in \XX$. The vector $z_v$ is known as the M\"{o}bius transform (or Harsanyi Dividend) of $v$. A cooperative game $v$ is \emph{totally monotone} when $z_v\ge 0$.\footnote{Totally monotone games are also known as `belief functions' in the theory of \citet{dempster1968generalization} and \citet{shafer1976mathematical}.} Any totally monotone game is convex.  We omit the proof of the following simple lemma.

\begin{lemma}\label{lemma-convex}
Suppose that $v$ is totally monotone (in particular, convex). Then $v$ is strictly convex if and only if $z_v(\{a,b\})>0$ for every pair of alternatives $a,b\in X$.
\end{lemma}

\section{Rationalizable marginals}\label{sec-consistent}
We start by characterizing when a marginal stochastic choice dataset $(\mu,\lambda)$ is generated by some conditional choice probabilities. We say that $(\mu,\lambda)$ is \emph{rationalizable} when it could have been generated by some stochastic choice function. Formally, 

\begin{definition}\label{def-consistent}
The marginal stochastic choice dataset $(\mu,\lambda) \in \Delta(\XX) \times \Delta(X)$ is rationalizable when there exists a collection $\pi=(\pi(\cdot|A))_{A\in \XX}$ with each $\pi(\cdot|A)\in \Delta(A)$, such that 
$$\lambda(a)= \sum_{\{A: a\in A\}} \mu(A)\pi(a|A)$$ 
for every $a\in X$. We say that $\pi$ rationalizes $(\mu,\lambda)$ when it satisfies the above condition. 
\end{definition}

Note that the definition places no restrictions on the form of the rationalizing stochastic choice function $\pi$. The only constraints are those of availability: If $a\notin A$, then we must have $\pi(a|A)=0$. It turns out that rationalizability can be characterized by a simple collection of inequalities reflecting these availability constraints. The following proposition has been obtained by various authors in different contexts, see for example \citet[Corollary 3]{chateauneuf1989some} and \citet[Theorem 7.20]{grabisch2016set}. This result serves as a useful benchmark for later sections where more restrictions are imposed on $\pi$. 

\begin{proposition}\label{prop-basic}
For a marginal stochastic choice dataset $(\mu,\lambda) \in \Delta(\XX) \times \Delta(X)$, the following are equivalent:
\begin{enumerate}
    \item $(\mu,\lambda)$ is rationalizable.
    
    \item For all $A \in \XX$,
    $ \sum_{ a \in A} \lambda(a) \ge \sum_{B \subseteq A} \mu(B).$
    
    \item For all $A \in \XX$, 
    $ \sum_{ a \in A} \lambda(a) \le \sum_{\{B:B\cap A\neq \emptyset\}} \mu(B).$ 
\end{enumerate}
\end{proposition}

Condition 2 of the proposition states that the total frequency with which elements of $A$ are chosen must be at least as large as the frequency with which \emph{only} alternatives from $A$ are available. Thus, this condition is clearly necessary for rationalizability. Similarly, condition 3 states that the frequency of choices from $A$ cannot exceed the frequency that \emph{some} alternative from $A$ is available. These two collections of inequalities are equivalent since the inequality in 2 for a menu $A$ is the same as the inequality in 3 for the menu $A^c$. The fact that these inequalities are sufficient for rationalizability can be proven using the max-flow min-cut duality theorem -- see Proposition \ref{prop-flow-cut} below for a more general result.\footnote{Proposition \ref{prop-basic} is a generalized version of the classic `marriage lemma' \citep{hall1934representation}. Theorem 3.1 in \citet{barseghyan2021heterogeneous} also applies a generalization of the marriage lemma to characterize the sharp identification region when the distribution of availability is unobservable but there is a known bound on the minimal size of menus.}

Recall from the previous section that for a given $\mu\in \Delta(\XX)$ we define the game $v_\mu$ by $v_\mu(A)= \sum_{B\subseteq A} \mu(B)$, and that the core of $v_\mu$ is the set of distributions in $\Delta(X)$ that are point-wise above $v_\mu$.\footnote{Since $\mu$ is a probability distribution the game $v_\mu$ is non-negative and therefore the core only contains non-negative vectors.} In other words, condition 2 of Proposition \ref{prop-basic} can be rewritten as $\lambda \in Core(v_\mu)$. We therefore have the following.

\begin{corollary}\label{coro-basic}
    The marginal stochastic choice dataset $(\mu,\lambda)$ is rationalizable if and only if $\lambda \in Core(v_\mu)$.
\end{corollary}

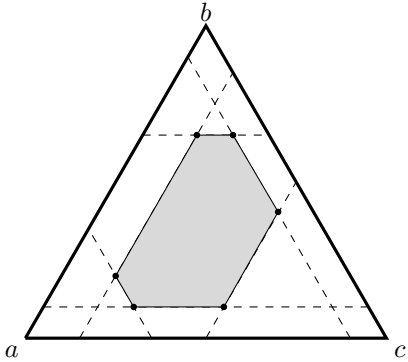
\begin{figure}
  \colorlet{GrayLight}{black!15}
  \colorlet{GrayMedium}{black!80}
  \begin{center}
    \begin{tikzpicture}[scale=1.2]
        \draw [very thick] (0,0) -- (4,0) -- (2,3.4641) -- (0,0);
        \node at (-0.15,-0.15) {$a$};
        \node at (4.15, -0.15) {$c$};
        \node at (2, 3.62) {$b$};
        
        \draw [dashed] (3.6,0) -- (1.8,3.1177);
        \draw [dashed] (1.4,0) -- (0.7,1.2124);
        
        \draw [dashed] (0.2,0.3464) -- (3.8,0.3464);
        \draw [dashed] (1.3,2.2517) -- (2.7,2.2517);
        
         \draw [dashed] (0.6,0) -- (2.3,2.9445);
        \draw [dashed] (2,0) -- (3,1.7321);
        
        \draw [fill=GrayLight] (1.2,0.3464) -- (2.199,0.3464) -- (2.8,1.4) -- (2.3,2.2517) -- (1.9,2.2517) -- (1,0.69) -- (1.2,0.3464);
        
        \draw [fill] (1.2,0.3464) circle [radius=0.03];
        \draw [fill] (2.199,0.3464) circle [radius=0.03];
        \draw [fill] (2.8,1.4) circle [radius=0.03];
        \draw [fill] (2.3,2.2517) circle [radius=0.03];
        \draw [fill] (1.9,2.2517) circle [radius=0.03];
        \draw [fill] (1,0.69) circle [radius=0.03];
    \end{tikzpicture}
  \end{center}
  \caption{The solid line outer triangle represents the set of choice distributions $\lambda$ over $X=\{a,b,c\}$. The shaded area is $Core(v_\mu)$ when $\mu$ is given by $\mu(a)=0.1$, $\mu(b)=0.1$, $\mu(c)=0.15$, $\mu(\{a,b\})=0.3$, $\mu(\{a,c\})=0.1$, $\mu(\{b,c\})=0.1$, and $\mu(\{a,b,c\})=0.1$. Each dashed line corresponds to a constraint $\sum_{a\in A} \lambda(a) \ge v_\mu(A)$ for some menu $A$.}
\label{figure-core}
\end{figure}

Figure \ref{figure-core} illustrates the shape of $Core(v_\mu)$ for a particular distribution $\mu$ in the case where $X$ contains three alternatives. By Corollary \ref{coro-basic} this is precisely the set of choice distributions $\lambda$ such that $(\mu,\lambda)$ is rationalizable. 

\medskip

It is worth pointing out that the rationalizing stochastic choice function $\pi$ is typically not unique whenever the number of alternatives in $X$ is at least 3. For a concrete example, let $X=\{a,b,c\}$ and let $\mu$ assign probability of $\frac{1}{3}$ to each of the binary menus. Consider a (deterministic) choice $\pi$ that selects $a$ from $\{a,b\}$, $b$ from $\{b,c\}$, and $c$ from $\{a,c\}$. This $\pi$ rationalizes the uniform distribution over $X$. However, $\pi'$ that chooses the other alternative in each menu also rationalizes the uniform distribution.




In what follows, we impose more structure on the stochastic choice function $\pi$ by considering some of the prominent models from the stochastic choice literature. We characterize the testable implications of each model for marginal stochastic choice probabilities and study what can be inferred about the model parameters.  


\section{Random utility}\label{sec-RUM}
The first model we consider is the random utility model (RUM). Suppose that each individual in a population has a strict preference ordering over $X$ and chooses their top alternative from the available menu. We examine whether the choices of such a population of rational decision makers puts additional testable restrictions on marginal stochastic choice data. We also examine what can be inferred about the distribution of preferences from observing the marginals.

Formally, let $O$ be the set of strict total orders on $X$, with typical element $\succ$. Given a menu $A$ and an alternative $a\in A$ denote $T[A,a] = \{\succ \in O ~:~ a\succ b ~~\forall b\in A\setminus\{a\} \}$ the collection of orders that rank $a$ above any other element in $A$.


\begin{definition}\label{def-RUM}
The marginal stochastic choice dataset $(\mu,\lambda)$ is RUM rationalizable when there is a distribution $\nu\in\Delta(O)$ such that 
$$\lambda(a)= \sum_{\{A: a\in A\}} \mu(A) \nu(T[A,a]) $$ for every $a\in X$. When $\nu$ satisfies the above equality, we say that $\nu$ RUM rationalizes the marginal stochastic choice dataset.
\end{definition}

The definition of RUM rationalizability is the same as saying that there is a distribution of conditional choice probabilities $\pi$ that rationalizes $(\mu,\lambda)$ with $\pi(a|A)=\nu(T[A,a])$. It is well-known \citep{Block1960, falmagne1978representation} that not every $\pi$ can be generated in this way. The novelty of our situation is that we only observe marginal frequencies, so it is unknown whether the random utility model adds any additional restrictions on top of unrestricted rationalizability. Namely, even when the marginal choice data $(\mu,\lambda)$ can be rationalized by a stochastic choice function $\pi$ inconsistent with RUM, it might be possible to find another stochastic choice function $\pi'$ that is consistent with RUM and rationalizes the same marginals. The next proposition says that this is indeed the case.

\begin{proposition}\label{prop-RUM}
The marginal stochastic choice dataset $(\mu,\lambda)$ is RUM rationalizable if and only if it rationalizable.
\end{proposition}

\begin{proof}
Clearly, we only need to show that if $(\mu,\lambda)$ is rationalizable, then it is RUM rationalizable. This is a simple consequence of the following result of 
\citet{Shapley1971}, which characterizes the extreme points of the core of a convex game.\footnote{See \citet[Remark 3.16]{grabisch2016set} for other papers that independently proved similar results.} Given $\succ\in O$ and an alternative $a$, denote by $L_\succ(a)=\{b: a\succ b\}$ the lower contour set of $a$ according to $\succ$. 
\begin{proposition}\label{prop-core-convex}
\citep{Shapley1971} Let $v$ be a convex game. Then $p\in \RR^X$ is an extreme point of $Core(v)$ if and only if there is $\succ\in O$ such that for every $a$ 
$$p(a)=v(L_\succ(a)\cup\{a\}) - v(L_\succ(a)).$$
Moreover, if $v$ is strictly convex then the mapping $\succ \longrightarrow p$ is one-to-one.
\end{proposition}

Now, suppose that $(\mu,\lambda)$ is rationalizable. Then $\lambda\in Core(v_\mu)$ by Proposition~\ref{prop-basic} and recall from Section~\ref{sec-preliminaries} that $v_\mu$ is totally monotone and hence convex. Since $Core(v_\mu)$ is a convex and compact polytope, $\lambda$ can be written as a convex combination of its extreme points. Thus, by Proposition~\ref{prop-core-convex} there is a distribution $\nu$ over $O$ such that for every $a\in X$,
$$\lambda(a) = \sum_{\succ\in O} \nu(\succ)[v_\mu(L_\succ(a)\cup\{a\}) - v_\mu(L_\succ(a))].$$

Now,
$$v_\mu(L_\succ(a)\cup\{a\}) - v_\mu(L_\succ(a)) = \sum_{\{A:~ a\in A \subseteq L_\succ(a)\cup \{a\}\}}\mu(A) = \sum_{\{A:~ a\in A,~ \succ\in T[A,a]\}}\mu(A),$$
where the first equality is by the definitions of $v_\mu$ and $L_\succ(a)$, and the second equality is by the definition of the set $T[A,a]$. Combining the above two equations, we get that
$$\lambda(a) = \sum_{\succ\in O} \nu(\succ) \sum_{\{A:~ a\in A,~ \succ\in T[A,a]\}}\mu(A) = \sum_{\{A: a\in A\}} \mu(A) \nu(T[A,a]),$$
where the last equality is just a change in the order of summation. 
\end{proof}

Before proceeding we note that Proposition \ref{prop-RUM} can also be deduced from the results of \citet{dempster1968upper} on belief functions and of \citet{weber1988probabilistic} on random order values. 

\medskip

It is typically impossible to back out the distribution over preferences $\nu$ from the marginal stochastic choice dataset $(\mu,\lambda)$. Indeed, it is well-known that point identification fails in the RUM even when conditional choice frequencies are observable.\footnote{\cite{barbera1986falmagne} and \citet{fishburn1998stochastic} give simple examples showing the non-uniqueness of the rationalizing distribution of preferences when there are four alternatives. These examples demonstrate that even the support of $\nu$ cannot be identified.} If $\nu$ and $\nu'$ generate the same stochastic choices in every menu, then clearly they cannot be separated based on the marginals. Of course, when only $(\mu,\lambda)$ is available the situation is even worse, since $\nu$ and $\nu'$ may not be distinguishable even when they are distinguishable based on conditional choice frequencies. We illustrate this with the following example.

\begin{example}\label{example-RUM}
Let $X=\{a,b,c\}$ and suppose that $\mu$ assign probability of $1/4$ to each of the menus $\{a,b\}$, $\{a,c\}$, $\{b,c\}$ and $X$. Let $\nu$ assign probability of $1/3$ to each of the orderings $a\succ b\succ c$, $b\succ c\succ a$, and $c\succ a \succ b$. Let $\nu'$ assign probability of $1/3$ to each of the other three orderings. Then $\nu$ and $\nu'$ induce different choice frequencies in each of the binary menus, but the resulting marginal distribution of choices $\lambda$ is uniform for both of them. 
\end{example}

The failure of identification can be easily understood by looking at Figure \ref{figure-core}. Recall that each of the six extreme points of $Core(v_\mu)$ corresponds to a homogeneous population of agents all having the same preferences. Any $\lambda$ in the interior of the core can be represented in multiple ways as a convex combination of these extreme points, and any such representation is a possible RUM rationalization of $(\mu,\lambda)$. More generally, for typical distributions $\mu$ the extreme points of $Core(v_\mu)$ are not affinely independent, so most $\lambda$'s have more than one representation as a convex combination of these vertices. For example, if $\mu$ has full support, then $v_\mu$ is strictly convex (Lemma \ref{lemma-convex}) and therefore $Core(v_\mu)$ has $n!$ different extreme points (Proposition~\ref{prop-core-convex}).




While point identification fails in general, it may be that some properties of the rationalizing distribution of preferences are identifiable. It is known for example that the probabilities of contour sets are pinned down by the (conditional) stochastic choices from menus \citep{falmagne1978representation, barbera1986falmagne}. This is no longer true when the dataset contains only the marginals $(\mu,\lambda)$. Indeed, in Example \ref{example-RUM} the probability that $a$ is ranked above $b$ and below $c$ is $1/3$ under $\nu$ but is $0$ under $\nu'$. However, we now show that it is still possible to test whether contour sets are not in the support of any RUM rationalization.

Given $\succ\in O$ and $A\in \XX$, we say that $A$ is $\succ$-inferior when $b\succ a$ for every $a\in A$ and $b\in A^c$. Also, for a distribution $\nu\in\Delta(O)$ say that $A$ is $\nu$-inferior when $A$ is $\succ$-inferior with $\nu$ probability one. In words, $A$ is $\nu$-inferior when every individual in the population ranks every element of $A$ below every element of $A^c$. The next proposition offers a simple test to determine whether a given set $A$ is inferior based only on the marginals. The proof is in the Appendix.

\begin{proposition}\label{prop-RUM-inferior}
Suppose the marginal stochastic choice dataset $(\mu,\lambda)$ is RUM rationalizable and that $\mu(\{a,b\})>0$ for every pair $a,b$. Fix $A\in \XX$. The following are equivalent:
\begin{enumerate}
\item $A$ is $\nu$-inferior for every $\nu\in \Delta(O)$ that RUM-rationalizes $(\mu,\lambda)$.
\item $A$ is $\nu$-inferior for some $\nu\in \Delta(O)$ that RUM-rationalizes $(\mu,\lambda)$.
\item $\sum_{a \in A} \lambda(a)=v_\mu(A)$.
\end{enumerate}
\end{proposition}

The following is a direct corollary of the last proposition.

\begin{corollary}
Under the assumptions of Proposition~\ref{prop-RUM-inferior}, the collection of menus satisfying $\lambda(A)=v_\mu(A)$ can be ordered so that $\emptyset\subset A_1 \subset A_2 \subset \ldots \subset A_M=X$. Furthermore, any $\succ$ in the support of a rationalizing $\nu$ satisfies $a_i\succ a_j$ whenever $i>j$, $a_j\in A_j$, and $a_i\in A_i\setminus A_j$.
\end{corollary}

While we have formulated the last results for the extreme case in which $A$ is inferior with $\nu$ probability one, it is also possible to use the marginals to obtain bounds on the probability that a menu $A$ is inferior or superior (meaning that $A^c$ is inferior). Specifically, it is not hard to show that under the assumptions of the proposition, the ratio $\frac{\lambda(A)-v_\mu(A)}{1-v_\mu(A)-v_\mu(A^c)}$ is an upper bound for the $\nu$ probability that $A$ is superior for every rationalizing $\nu$. This ratio is also a lower bound for the $\nu$ probability that $A$ is \emph{not} inferior for every rationalizing $\nu$. 

Finally, one may want to know when is it the case that $(\mu,\lambda)$ does have a unique rationalizing $\nu$.\footnote{\cite{turansick2022identification} characterizes those conditional stochastic choice functions that admit a unique representation, and shows that this is equivalent to the support of the representation being unique.} Under the assumptions of Proposition \ref{prop-RUM-inferior}, this happens only when $\lambda$ is in one of the edges (one-dimensional faces) of $Core(v_\mu)$. Equivalently, $(\mu,\lambda)$ has a unique RUM rationalization if and only if for every $k=2,\ldots,n$ there is a menu $A$ of cardinality $k$ such that $\lambda(A)=v_\mu(A)$.


\section{The Luce model}\label{sec-Luce}

The following definition formalizes the idea that the marginal stochastic choice dataset $(\mu,\lambda)$ is consistent with an individual who behaves according to the Luce model of stochastic choice.

\begin{definition}
The marginal stochastic choice dataset $(\mu,\lambda)$ is Luce rationalizable when there is $u\in Int(\Delta(X))$ such that for every $a\in X$ 
\begin{equation}\label{eqn-Luce}
    \lambda(a)= \sum_{\{A: a\in A\}} \mu(A) \frac{u(a)}{\sum_{b \in A} u(b)}.
\end{equation} 
We say $u$ Luce rationalizes $(\mu,\lambda)$ when (\ref{eqn-Luce}) holds.\footnote{We assume that the sum of coordinates of $u$ is 1. Clearly, any positive vector can be normalized without changing the resulting stochastic choice function.}
\end{definition}

The Luce model is a special case of the RUM. Nevertheless, in the next proposition we show that RUM and Luce rationalizations are essentially the same in terms of the marginals that they can generate. Under a mild assumption on the marginal distribution over menus ($\mu$), any `interior' marginal distribution of choices ($\lambda$) that is RUM rationalized can also be Luce rationalized. However, in contrast to RUM rationalizability, under Luce rationalizability we can identify the values of $u$ and therefore deduce the conditional choice probabilities in every menu by observing only the marginals.

\begin{proposition}\label{prop-homeomorphism}
Suppose that for every pair of alternatives $a,b\in X$ there is a menu $A$ in the support of $\mu$ such that $\{a,b\}\subseteq A$. Then the mapping $u\longrightarrow \lambda$ given by (\ref{eqn-Luce}) is a bijection between $Int(\Delta(X))$ and $Int(Core(v_\mu))$.
\end{proposition}

The proof of Proposition \ref{prop-homeomorphism} can be found in the Appendix. Recall that $Core(v_\mu)$ is exactly the set of $\lambda$'s that can be rationalized by some (unrestricted) stochastic choice function $\pi$, as well as those that can be rationalized by the RUM. The proposition implies that if the constraints hold strictly at $\lambda$, meaning $\lambda(A)>v_\mu(A)$ for all menus $A\in \XX \setminus \{X\}$, then $\lambda$ can also be rationalized by the Luce model. Distributions of choices $\lambda$ on the relative boundary cannot be rationalized with the Luce model because the weights $u$ must be strictly positive. The assumption that every pair of alternatives is contained in some menu in the support of $\mu$ guarantees that $Core(v_\mu)$ has the same dimension as the entire simplex $\Delta(X)$. In fact, the proof of Proposition~\ref{prop-homeomorphism} shows that the mapping $u\longrightarrow \lambda$ given by (\ref{eqn-Luce}) is a homeomorphism from $Int(\Delta(X))$ to $Int(Core(v_\mu))$.

In cooperative game theory, the distribution $\lambda$ obtained from the game $v_\mu$ via equation (\ref{eqn-Luce}) is known as a positively weighted Shapley value, where the weights are given by the vector $u$. When $u$ is the constant vector one obtains the standard Shapley value of the game $v_\mu$. \citet{monderer1992weighted} prove almost the same result as Proposition \ref{prop-homeomorphism} above, showing that for a strictly convex game $v$ the core is homeomorphic to the set of all weighted Shapley values (allowing for lexicographic systems of weights). Our result is weaker in that we only consider positive weights and the games we consider are totally monotone which implies convexity. On the other hand, our assumption on the support of $\mu$ is weaker than strict convexity of $v_\mu$ (recall Lemma \ref{lemma-convex}). Our proof is also somewhat simpler due to the total monotonicity of $v_\mu$.

\medskip

Since under the Luce model conditional choice probabilities are pinned down by the marginals, it is natural to ask whether and when the two kinds of distributions share similar properties. The next result gives an example where this is the case. We show that if the marginal distribution of menu availability ($\mu$) satisfies a certain symmetry property between two alternatives, then the alternative that has higher marginal choice probability also has a higher conditional choice probability in every menu that contains both.  

\begin{definition}\label{def-exchangeable}
For $a,b\in X$, the marginal distribution of menu availability $\mu\in \Delta(\XX)$ is $ab$-exchangeable when $\mu(A\cup \{a\}) = \mu(A\cup \{b\})$ for every $A\subseteq X\setminus\{a,b\}$. The distribution $\mu$ is exchangeable when it is $ab$-exchangeable for every pair $a,b\in X$. 
\end{definition}

\begin{proposition}\label{prop-Luce-symmetry}
Suppose that $u$ Luce rationalizes $(\mu, \lambda)$ and that $\mu$ is $ab$-exchangeable. Then $\lambda(a)\ge \lambda(b)$ if and only if $u(a)\ge u(b)$. In particular, when $\mu$ is exchangeable then the ranking of menu-contingent choice probabilities is the same as the ranking of marginal choice probabilities.
\end{proposition}

The proof of Proposition \ref{prop-Luce-symmetry} is in the Appendix.


\section{Random consideration sets}\label{sec-RCS}

We next consider the independent random consideration set (IRCS) model of \citet{manzini2014stochastic}. For the IRCS model, a decision maker has a single preference ordering over alternatives and each alternative has a fixed menu-independent probability of being considered. When the menu $A$ is available, the probability that $a\in A$ is chosen is equal to the probability that $a$ is considered and that all alternatives in $A$ that the decision maker prefers to $a$ are not considered. 

In order to analyze the IRCS model, we need to slightly modify our framework to account for the possibility that the decision maker does not consider any of the available alternatives. As in \cite{manzini2014stochastic}, we add an outside option that is always available and is chosen only when no other available alternative is considered.

Formally, let $X^*=X\cup\{x^*\}$, where $X$ is the original set of alternatives as before and $x^*\notin X$ is an outside option. For each menu $A\in \XX$ we denote $A^*=A\cup\{x^*\}$, and we let $\XX^*=\{A^*: A\in \XX\}$ be the collection of observed menus. This assumes that at least one of the original alternatives is available in any menu, though it would not be hard to relax this assumption.

Slightly abusing notation, we let the marginal distribution of menu availability be $\mu\in \Delta(\XX^*)$ and the marginal distribution of choices be $\lambda\in \Delta(X^*)$, so that $(\mu,\lambda)$ is the observable marginal stochastic choice dataset. The following definition formalizes consistency with the IRCS model.

\begin{definition}\label{def-RCS}
For a given $\succ\in O$, the marginal stochastic choice dataset $(\mu,\lambda)$ is $\succ$-IRCS rationalizable when there is $\gamma=(\gamma_a)\in [0,1]^X$ such that 
\begin{equation}\label{eqn-RCS}
    \lambda(a) = \sum_{\{A^*\in \XX^*:~ a\in A^*\}} \mu(A^*) \left[\gamma_a\prod_{\{b\in A:~ b\succ a\}}(1-\gamma_b)\right]
\end{equation}
for every $a\in X$. The marginal stochastic choice dataset $(\mu,\lambda)$ is IRCS rationalizable when it is $\succ$-IRCS rationalizable for some $\succ\in O$.
\end{definition}

When $(\mu,\lambda)$ is IRCS rationalizable and $(\succ, \gamma)$ satisfies equation (\ref{eqn-RCS}) we say that $(\succ,\gamma)$ IRCS rationalizes $(\mu,\lambda)$; when the order $\succ$ is given then we instead say that $\gamma$ $\succ$-IRCS rationalizes $(\mu,\lambda)$.

Note that the above definition does not specify the probability $\lambda(x^*)$ that the outside option is selected. This probability corresponds to the event that no other alternative is selected, meaning $\lambda(x^*)= 1-\sum_{a\in X}\lambda(a)$. It is easy to see that this is equal to the probability that none of the alternatives in $X$ are considered, which happens with probability $\prod_{b\in A}(1-\gamma_b)$ when $A^*$ is the realized menu. 

The next proposition takes the ordering $\succ$ as given and characterizes when the marginals $(\mu,\lambda)$ are $\succ$-IRCS rationalizable.

\begin{proposition}\label{prop-RCS}
For the marginal stochastic choice dataset $(\mu,\lambda)$, assume that $\mu(\{a,x^*\})>0$ for all $a\in X$. Let $\succ\in O$, index the alternatives in $X$ so that $a_1\succ\ldots\succ a_n$, and denote $A_k=\{a_1,\ldots,a_k\}$ for $1\le k\le n$. Define
$$t^\succ_1 = \lambda(a_1)\left[\mu(\{A^*\in\XX^*:~ a_1\in A^*\})\right]^{-1},$$
and for each $k=2,\ldots,n$ define recursively 
$$t^\succ_k=\lambda(a_k)\left[ \sum_{B\subseteq A_{k-1}} \mu(\{A^*\in \XX^*:~ A^*\cap A_k=B\cup\{a_k\}\}) \prod_{a_i\in B}(1-t^\succ_i) \right]^{-1}.$$
The dataset $(\mu,\lambda)$ is $\succ$-IRCS rationalizable if and only if $t^\succ_k \le 1$ for every $1\le k\le n$. Furthermore, if $(\mu,\lambda)$ is $\succ$-IRCS rationalizable then letting $\gamma_{a_k}=t^\succ_k$ uniquely $\succ$-IRCS rationalizes $(\mu,\lambda)$.
\end{proposition}

The proof of Proposition \ref{prop-RCS} is in the Appendix. The assumption that $\mu(\{a,x^*\})>0$ for all $a\in X$ guarantees that each $t^\succ_k$ is well-defined. This result demonstrates how to test for consistency with the IRCS model for a given preference relation and how to determine the consideration probabilities $\gamma$ from $(\mu,\lambda)$. The numbers $t^\succ_k$ can be easily calculated from the data, starting from $t^\succ_1$ for the most preferred alternative and continuing recursively in decreasing order. If any of the $t^\succ_k$'s exceeds one, then $\succ$-IRCS is ruled out. Otherwise, if all $t^{\succ}_k \le 1$, then $\gamma_k=t^\succ_k$ uniquely rationalizes the data for the preference relation $\succ$. 

Assuming that the ranking $\succ$ of the agent is known seems unreasonable for many applications. To test whether $(\mu,\lambda)$ is IRCS rationalizable one would need to calculate the vector $t^\succ$ for each $\succ\in O$. From Proposition~\ref{prop-RCS} we have the following immediate corollary. 

\begin{corollary}\label{coro-RCS1}
Under the conditions of Proposition \ref{prop-RCS}, the marginal stochastic choice dataset  $(\mu,\lambda)$ is IRCS-rationalizable if and only if there is $\succ\in O$ such that $t^\succ_k\le 1$ for every $1\le k\le n$.
\end{corollary}

We note that while the consideration probabilities are uniquely identified for a given preference order, a dataset can sometimes be IRCS-rationalized by several preference orders. Indeed, in general there may be $(\succ,\gamma)$ and $(\succ',\gamma')$ that both IRCS-rationalize $(\mu,\lambda)$.\footnote{This stands in stark contrast to the case where the dataset contains stochastic choices from a rich collection of menus, as shown in \citet[Theorem 1]{manzini2014stochastic}.} We illustrate this with the following example.

\begin{example}\label{example-RCS}
Let $X=\{a,b\}$ and suppose that $\mu$ assigns probability of 1/3 to each of the menus $\{a,x^*\},$ $\{b,x^*\}$ and $\{a,b,x^*\}$. Consider first the ordering $a\succ b$. It is easy to check using Proposition \ref{prop-RCS} that $(\mu, \lambda)$ is $\succ$-IRCS rationalizable if and only if $0\le \lambda(a)\le 2/3$ and $0\le \lambda(b)\le 2/3-0.5\lambda(a)$. Similarly, for the ordering $b\succ' a$ we have that $(\mu, \lambda)$ is $\succ'$-IRCS rationalizable if and only if $0\le \lambda(b)\le 2/3$ and $0\le \lambda(a)\le 2/3-0.5\lambda(b)$. 

Now, consider $\lambda(a)=\lambda(b)=1/3$ (and $\lambda(x^*)=1/3$). This can be rationalized either by the ordering $\succ$ with $\gamma_a=0.5$ and $\gamma_b=2/3$, or by the ordering $\succ'$ with $\gamma'_b=0.5$ and $\gamma'_a=2/3$. Thus, the ordering and consideration probabilities cannot be identified from $(\mu,\lambda)$.
\end{example}

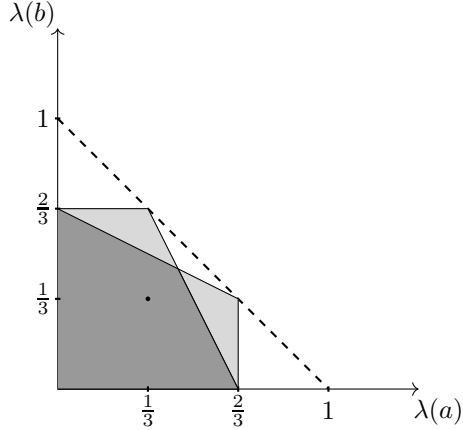
\begin{figure}
  \colorlet{GrayLight}{black!15}
  \colorlet{GrayMedium}{black!40}
  \begin{center}
    \begin{tikzpicture}[scale=1.2]
      \draw [->] (0, 0) -- (4, 0);
      \draw [->] (0, 0) -- (0, 4);
      \node at (-0.28, 4.15) {$\lambda(b)$};
      \node at (4.22, -0.25) {$\lambda(a)$};
      \node at (1, -0.24) {$\frac{1}{3}$};
      \node at (2, -0.24) {$\frac{2}{3}$};
      \node at (3, -0.24) {$1$};
      \node at (-0.16, 1) {$\frac{1}{3}$};
      \node at (-0.16, 2) {$\frac{2}{3}$};
      \node at (-0.16, 3) {$1$};

      \draw [thick, dashed] (0, 3) -- (3, 0);
      
      \draw [fill=GrayLight] (0,0) -- (2,0) -- (2,1) -- (0,2);
      \draw [fill=GrayLight] (0,0) -- (0,2) -- (1,2) -- (2,0);
      \draw [fill=GrayMedium] (0,0) -- (2,0) -- (1.333,1.333) -- (0,2);
      
      \draw [fill] (1,1) circle [radius=0.02];
      
      \draw [thick] (1, -.03) -- (1, .03);
      \draw [thick] (2, -.03) -- (2, .03);
      \draw [thick] (3, -.03) -- (3, .03);
      \draw [thick] (-.03, 1) -- (.03, 1);
      \draw [thick] (-.03, 2) -- (.03, 2);
      \draw [thick] (-.03, 3) -- (.03, 3);
      
    \end{tikzpicture}
  \end{center}
  \caption{The figure illustrates the set of choice distributions rationalizable by the IRCS model from Example \ref{example-RCS}. The dark-shaded area (with the marked point $\lambda(a)=\lambda(b)=\frac{1}{3}$) contains $\lambda$'s that can be rationalized by both orders $a\succ b$ and $b\succ' a$. Each of the light-shaded areas can be rationalized by only one of these orderings.}\label{figure-RCS}
\end{figure}

Figure \ref{figure-RCS} illustrates the non-identifiability problem in Example \ref{example-RCS}. The dark-shaded region, which includes the point $\lambda(a)=\lambda(b)=\frac{1}{3}$ from the example, contains choice distributions that can be rationalized by both orders $a\succ b$ and $b\succ' a$. 
Notice however that when $\lambda(a)+\lambda(b)$ is close enough to 1, i.e., when the outside option is only chosen rarely, the rationalizing order is unique. In the next corollary we show that this is always the case: If the outside option is selected with small enough probability, then both $\succ$ and $\gamma$ can be identified.

\begin{corollary}\label{coro-RCS2}
Assume that $\mu\in \Delta(\XX^*)$ has full support. There is a constant $\bar{\delta}>0$ such that if $\lambda\in \Delta(X^*)$ satisfies $\lambda(x^*)<\bar\delta$, and if $(\mu,\lambda)$ is IRCS-rationalizable, then there is a unique $(\succ, \gamma)$ that IRCS-rationalizes $(\mu,\lambda)$.
\end{corollary}

The proof of Corollary \ref{coro-RCS2} is in the Appendix.

\section{Endogenous availability}\label{sec-2-period}

In the models considered up to now the marginal distribution of menu availability $\mu$ was viewed as exogenously given and choices made by the agent (or population of agents) were only affecting the marginal choice probabilities $\lambda$. We now consider a conceptually different scenario in which the agent first chooses a menu of alternatives and then chooses an alternative from that menu in the second period. The choice of a menu in the first period not only determines what alternatives are available in the second period, but also implies something about the agent's preferences over alternatives, and therefore also about the alternative that the agent eventually chooses. Thus, unlike models such as RUM where the distribution of preferences over alternatives is independent of the menu, for the models considered below preferences of individuals who choose different menus can be different.

We analyze two prominent models of preferences over menus that represent two polar considerations. The first is based on \citet{gul2001temptation}, where preferences over menus are affected by anticipated temptation in the second stage. Here agents are worried about lack of self-control and therefore prefer to limit their future choices. The second model is in the spirit of \citet{kreps1979representation}. It describes agents who are unsure about their future tastes and therefore have preference for flexibility. We characterize when the observed marginal stochastic choice dataset $(\mu,\lambda)$ can be described by a population of agents who have preferences described by each of these models.

\subsection{Temptation and self-control}\label{subsec-TSC}

We start with the model of \citet{gul2001temptation}. An agent is characterized by a pair of functions $u,v:X\to \RR$. The function $u$ describes the agent's value for the alternatives when temptation is absent, while the function $v$ can be interpreted as the agent's urges when choosing an alternative from a menu. Given $(u,v)$, preferences over menus are represented by the utility function 
$$U_{u,v}(A) = \max_{a\in A}[u(a)+v(a)]-\max_{b\in A}[v(b)].$$
After choosing a menu, the agent chooses an alternative from that menu that maximizes the sum $u(a)+v(a)$. 

The following definition formalizes the idea that the marginals $(\mu,\lambda)$ are consistent with a population of individuals who behave according to the above temptation and self-control model.  

\begin{definition}\label{def-gul-pese}
Let $\XX'\subseteq \XX$ be a non-empty collection of feasible menus. The marginal stochastic choice dataset $(\mu,\lambda)$ is temptation and self-control rationalizable given $\XX'$ ($\XX'$-TSC rationalizable, for short) when there is a distribution $\psi$ over $\RR^X\times \RR^X$ such that:\\
(i) For every $A\in \XX'$, 
$$\mu(A) = \psi\Big(\big\{(u,v):~ U_{u,v}(A)>U_{u,v}(B)~~ \forall B\in \XX'\setminus \{A\} \big\}\Big).$$
(ii) For every $a\in X$,
\begin{eqnarray*}
\lambda(a) = \sum_{\{A\in \XX': a\in A\}} \psi \Big(\big\{ (u,v) & : & u(a)+v(a)>u(b)+v(b)~~ \forall b\in A\setminus\{a\} \textit{ and }\\
& & U_{u,v}(A)>U_{u,v}(B)~~ \forall B\in \XX'\setminus \{A\}\big\}\Big).
\end{eqnarray*}
\end{definition}

Condition (i) from Definition~\ref{def-gul-pese} means that the observed marginal distribution over menus matches the population preference distribution over menus, while condition (ii) is the analogous condition for alternatives. Note that we require strict inequalities at both stages so that individuals strictly prefer their chosen menu over any other feasible menu, and that they strictly prefer their chosen alternative over any other alternative in the menu they selected. Allowing for ties both complicates the notation and trivializes the problem. 

We also note that consistency of a temptation and self control model with marginal stochastic choice data depends on the collection of feasible menus $\XX'$. If $\XX'$ contains every menu ($\XX'=\XX$), then every individual would choose a singleton menu in the first stage, and the problem becomes uninteresting. Restricted domains of feasible menus are natural in many applications and are often used in the stochastic choice literature.   

To characterize $\XX'$-TSC rationalizability, we need to introduce some additional notation. First, for any feasible menu $A\in \XX'$ let $\bar{A}_{\XX'} = A\setminus \bigcup_{\{B\in\XX':~ B\subsetneq A\}} B$ be the set of alternatives in $A$ that are not contained in any feasible sub-menu of $A$. We say that $A\in\XX'$ is redundant when $\bar A_{\XX'}=\emptyset$ and that $A$ is non-redundant otherwise. The collections of redundant and non-redundant menus in $\XX'$ are denoted by $R(\XX')$ and $NR(\XX')$, respectively. Intuitively, an agent who chooses $A$ in the first period must choose an alternative from  $\bar A_{\XX'}$ in the second period, since if she was interested in one of the other alternatives of $A$ then she could have chosen a smaller menu in the first period and reduce the temptation. In particular, a redundant menu should never be chosen under the TSC model.

Second, given $\mu\in\Delta(\XX')$ we define a game $v^{\XX'}_\mu$ by $v^{\XX'}_\mu(A) = \sum_{\{B\in \XX' : \bar{B}_{\XX'} \subseteq A \}} \mu(B)$ for every non-empty $A\subseteq X$. The definition of $v^{\XX'}_\mu$ differs from that of $v_\mu$ of Section \ref{sec-consistent} in that here we sum up the probabilities of all menus $B$ such that $\bar B_{\XX'} \subseteq A$ rather than only menus $B$ that are contained in $A$ themselves. Thus, we clearly have that $v^{\XX'}_\mu \ge v_\mu$.

\begin{proposition}\label{prop-TSC}
The marginal stochastic choice dataset $(\mu,\lambda)$ is $\XX'$-TSC rationalizable if and only if $\mu(R(\XX'))=0$ and $\lambda \in Core(v^{\XX'}_\mu)$.
\end{proposition}

Compared with Proposition~\ref{prop-basic}, the conditions for TSC rationalizability are stronger in two ways than the conditions for (unrestricted) rationalizability. The first is that redundant menus cannot be in the support of $\mu$, and the second is that the core constraints are calculated based on the game $v^{\XX'}_\mu$ which as explained above is larger than the game  $v_\mu$. We illustrate the proposition with the following example.

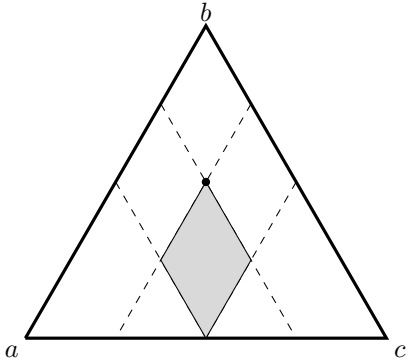
\begin{figure}
  \colorlet{GrayLight}{black!15}
  \colorlet{GrayMedium}{black!80}
  \begin{center}
    \begin{tikzpicture}[scale=1.2]
        \draw [very thick] (0,0) -- (4,0) -- (2,3.4641) -- (0,0);
        \node at (-0.15,-0.15) {$a$};
        \node at (4.15, -0.15) {$c$};
        \node at (2, 3.62) {$b$};
        
        \draw [dashed] (1,1.732) -- (2,0);
        \draw [dashed] (3,1.732) -- (2,0);
        \draw [dashed] (1.5,2.598) -- (3,0);
        \draw [dashed] (2.5,2.598) -- (1,0);
        
        \draw [fill=GrayLight] (2,0) -- (1.5,0.866) -- (2,1.732) -- (2.5,0.866) -- (2,0);
        
        \draw [fill] (2,1.732) circle [radius=0.04];
    \end{tikzpicture}
  \end{center}
  \caption{An illustration of Example \ref{example-TSC} in the case where $\mu(\{a\}) = \mu(\{c\}) = \mu(\{a,b\}) = \mu(\{b,c\}) = \frac{1}{4}$. The shaded area contains all choice distributions $\lambda$ such that $(\mu,\lambda)$ is rationalizable (the core of $v_\mu$). The only $\lambda$ such that $(\mu,\lambda)$ is $\XX'$-TSC rationalizable is $\lambda(a)=\lambda(c)=\frac{1}{4}$ and $\lambda(b)=\frac{1}{2}$ -- the marked point in the figure.}\label{figure-TSC}
\end{figure}

\begin{example}\label{example-TSC}
Let $X=\{a,b,c\}$ and $\XX'=\{\{a\}, \{c\}, \{a,b\}, \{b,c\}, X\}$. The only redundant menu in $\XX'$ is $X$, that is $R(\XX')=\{X\}$, so $\XX'$-TSC rationalizability requires that $X$ is not in the support of $\mu$. The second condition, $\lambda \in Core(v^{\XX'}_\mu)$, requires that $\lambda(a)\ge \mu(a)$, $\lambda(c)\ge \mu(c)$, as well as $\lambda(b)\ge \mu(\{a,b\})+\mu(\{b,c\})$.\footnote{The inequalities corresponding to other sets $A\subseteq X$ are implied by those for singleton menus in this example, but this is not the case  in general.} The latter inequality is due to 
$\overline{\{a,b\}}_{\XX'} = \overline{\{b,c\}}_{\XX'} =\{b\}$, which implies that $v_\mu^{\XX'}(\{b\}) = \mu(\{a,b\})+\mu(\{b,c\})$. Intuitively, agents who chose one of these two menus in the first stage necessarily choose alternative $b$ in the second stage, since if they like either $a$ or $c$ better than $b$ then they would have chosen these singleton menus to avoid the temptation. It follows that for every $\mu$ there is a unique $\lambda$ such that $(\mu,\lambda)$ is $\XX'$-TSC rationalizable, namely, $\lambda(a)= \mu(a)$, $\lambda(c)= \mu(c)$, and $\lambda(b)= \mu(\{a,b\})+\mu(\{b,c\})$. 

In contrast, if any stochastic choice function $\pi$ is allowed, then the inequalities that characterize rationalizability are given by $\lambda(a)\ge \mu(a)$, $\lambda(c)\ge \mu(c)$, $\lambda(a)+\lambda(b)\ge \mu(a)+\mu(\{a,b\})$, and $\lambda(b)+\lambda(c)\ge \mu(c)+\mu(\{b,c\})$. Thus, for typical distributions $\mu$ there would be many different $\lambda$'s such that $(\mu,\lambda)$ is rationalizable.

Figure~\ref{figure-TSC} illustrates the difference between (unrestricted) rationalizability and $\XX'$-TSC rationalizability for this example in the case where $\mu(\{a\}) = \mu(\{c\}) = \mu(\{a,b\}) = \mu(\{b,c\}) = \frac{1}{4}$. 
\end{example}

We now turn to the proof of the proposition. The main steps of the proof are outlined below, while more technical details are relegated to the Appendix.

\begin{proof}[Proof of Proposition~\ref{prop-TSC}]
Since the collection of feasible menus $\XX'$ is fixed, we sometimes omit it from the notation and write $\bar{A}$, $R$, and $NR$ instead of $\bar A_{\XX'}$, $R(\XX')$, and $NR(\XX')$, respectively. We still write $v^{\XX'}_\mu$ to avoid confusion with the game $v_\mu$ as defined earlier in the paper.

The first step is the following lemma. Its proof is in the Appendix.
\begin{lemma}\label{lemma-TSC-1}
Fix $A\in \XX'$ and $a\in A$. There exists $(u,v)\in \RR^X \times \RR^X$ such that $U_{u,v}(A)>U_{u,v}(B)$ for all $B\in \XX'\setminus\{A\}$ and such that $u(a)+v(a)>u(b)+v(b)$ for all $b\in A\setminus\{a\}$ if and only if $a\in \bar A$. Moreover, if $A\in R(\XX')$ then there is no $(u,v)\in \RR^X \times \RR^X$ such that $U_{u,v}(A)>U_{u,v}(B)$ for all $B\in \XX'\setminus\{A\}$.
\end{lemma}

The next lemma, which easily follows from the previous one, argues that for $(\mu,\lambda)$ to be $\XX'$-TSC rationalizable it is necessary and sufficient to consider only non-redundant menus and only alternatives that cannot be found in sub-menus. A proof is in the Appendix.

\begin{lemma}\label{lemma-TSC-2}
The pair $(\mu,\lambda)$ is $\XX'$-TSC rationalizable if and only if there exists a collection $\pi=\{\pi(\cdot|A)\}_{A\in NR}$, such that the support of each $\pi(\cdot|A)$ is contained in $\bar A$, and such that for every $a\in X$
$$\lambda(a)= \sum_{A\in NR} \mu(A)\pi(a|A).$$  
\end{lemma}

The next step is the heart of the proof, so we state it as a separate proposition. Its proof is based on the max-flow min-cut duality theorem and can be found in the Appendix.\footnote{Proposition \ref{prop-flow-cut} implies Proposition \ref{prop-basic} of Section \ref{sec-consistent} as a special case. Indeed, just take $Y=\XX$ and $h(A)=A$ for every $A\in \XX$. The characterizing condition becomes $\lambda\in Core(v_\mu)$.}

\begin{proposition}\label{prop-flow-cut}
Let $Y$ be a finite set, $\mu\in \Delta(Y)$, $\lambda\in \Delta(X)$, and fix a mapping $h:Y\to \XX$ where $\XX=2^X \setminus \emptyset$. There is a conditional probability system  $\pi=\{\pi(\cdot | y)\in \Delta(h(y))\}_{y \in Y}$ such that 
$\lambda(a) = \sum_{y \in Y} \mu(y)\pi(a|y)$ for every $a\in X$ if and only if for every $A\in \XX$ $$\sum_{a \in A} \lambda(a) \ge \sum_{\{y\in Y :~h(y)\subseteq A\}}\mu(y).$$
\end{proposition}

We can now finish the proof of the proposition.

\medskip

\noindent `Only If':
Suppose that $(\mu,\lambda)$ is $\XX'$-TSC rationalizable. By Lemma \ref{lemma-TSC-1}, we have that $\mu(R(\XX'))=0$, which proves the first condition of the proposition. Next, let $\pi$ be as in Lemma \ref{lemma-TSC-2}. Since the support of $\mu$ is contained in $NR$ we can apply Proposition \ref{prop-flow-cut} with $Y=NR$ and $h(B)=\bar B$ for every $B\in NR$. We thus get that, for every $A\in \XX$,
$$\sum_{a \in A} \lambda(a) \ge \sum_{\{B\in NR:~ \bar B\subseteq A\}} \mu(B) = \sum_{\{B\in \XX':~ \bar B\subseteq A\}}\mu(B) = v^{\XX'}_\mu(A),$$
where the inequality is from Proposition \ref{prop-flow-cut} , the first equality follows from $\mu(R(\XX'))=0$, and the last equality from the definition of the game $v^{\XX'}_\mu$. It follows that $\lambda\in Core(v^{\XX'}_\mu)$ as needed.

\medskip

\noindent `If':
Suppose that the conditions of the proposition hold. Applying Proposition \ref{prop-flow-cut} with $Y=NR$ and $h(A)=\bar A$ for every $A\in NR$ we get that there exists $\pi=\{\pi(\cdot | A)\in \Delta(\bar A)\}_{A\in NR}$ such that $\lambda(a) = \sum_{A\in NR} \mu(A)\pi(a|A)$. By Lemma \ref{lemma-TSC-2},  $(\mu,\lambda)$ is $\XX'$-TSC rationalizable and the proof is complete.
 
\end{proof}

\subsection{Preference for flexibility}\label{subsec-flexibility}

In the model of \citet{kreps1979representation}, when agents choose a menu in the first period they are still uncertain about the utility of each alternative. This uncertainty is captured by a distribution $\theta$ over $\RR^X$ describing the likelihood of realizations of utilities.\footnote{\citet{kreps1979representation} makes the state space of uncertainty explicit, but here it is more convenient to work directly with the distribution of utilities.} Given $\theta$, let $(Z_a)_{a\in X}$ be real-valued random variables whose joint distribution is $\theta$. Preferences over menus are represented by the function
$$U_\theta(A) = \EE\left(\max_{a\in A} Z_a\right).$$
In the second period the uncertainty is resolved and a vector of utilities $(z_a)_{a\in X}$ is realized. The agent chooses the alternative $a$ from the menu $A$ selected in the first period for which $z_a$ is maximal. 

Given a collection of feasible menus $\XX'\subseteq \XX$ and a menu $A\in \XX'$ we let
$$\Theta_{\XX'}(A) = \big\{\theta:~ U_\theta(A)>U_\theta(B)~~ \forall B\in \XX'\setminus \{A\} \big\}$$
be the set of distributions $\theta$ for which $A$ is the (unique) optimal menu among $\XX'$.
 
\begin{definition}\label{def-kreps}
Let $\XX'\subseteq \XX$ be a non-empty collection of feasible menus. The marginal stochastic choice dataset $(\mu,\lambda)$ is rationalized by a  preference for flexibility given $\XX'$, a $\XX'$-PF rationalization for short, when there is a distribution $\psi$ over\footnote{We slightly abuse notation and write $\Delta\left(\RR^X\right)$ for the space of probability distributions over the Borel sets of $\RR^X$. Since $\psi$ is a distribution over $\Delta\left(\RR^X\right)$, we need to specify the $\sigma$-algebra of measurable sets in this space. One way to do this is to endow $\Delta\left(\RR^X\right)$ with the total variation metric and use the corresponding Borel $\sigma$-algebra. Note however that in our sufficiency proof of Proposition \ref{prop-kreps} below we only use distributions $\psi$ that have finite support.} $\Delta\left(\RR^X\right)$ such that:\\
(i) For every $A\in \XX'$, 
$$\mu(A) = \psi\Big(\Theta_{\XX'}(A)\Big).$$
(ii) For every $a\in X$,
\begin{eqnarray*}
\lambda(a) = \sum_{\{A\in \XX': a\in A\}} ~~ \int\displaylimits_{\Theta_{\XX'}(A)} \theta \Big( \big\{ z\in \RR^X :~ z_a>z_b~~ \forall b\in A\setminus\{a\} \big\}\Big) \mathrm{d}\psi(\theta).
\end{eqnarray*}
\end{definition}

Condition (i) of Definition~\ref{def-kreps} guarantees that $\mu$ matches the preferences over menus in the population. To understand condition (ii), note that the integrand is the probability that an agent with distribution $\theta$ would choose $a$ in the second period conditional on choosing menu $A$ in the first period. Thus, the integral is the probability that an agent who chooses the menu $A$ ends up choosing the $a$th alternative. Summing up over all menus that contain $a$ then gives the overall frequency of $a$ in the population. Note that we require the optimum to be unique in both stages.

The characterization of $\XX'$-PF rationalizability turns out to be somewhat simpler than in the temptation and self control model. The only new substantial restriction added by endogenous availability is that agents never choose a menu that is contained in another feasible menu. We let 
$$\tilde R(\XX')=\{A\in \XX' ~:~ A\subsetneq B \textit{ for some } B\in \XX'\}$$ be those menus that are contained in another feasible menu. We have the following. 

\begin{proposition}\label{prop-kreps}
If the stochastic marginal choice dataset $(\mu,\lambda)$ is $\XX'$-PF rationalizable, then $\mu(\tilde R(\XX'))=0$ and $\lambda \in Core(v_\mu)$. Conversely, if $\mu(\tilde R(\XX'))=0$ and $\lambda \in Int(Core(v_\mu))$, then $(\mu,\lambda)$ is $\XX'$-PF rationalizable.
\end{proposition}

The proof of this proposition is in the Appendix. The small gap between the necessary condition $\lambda \in Core(v_\mu)$ and the sufficient condition $\lambda \in Int(Core(v_\mu))$ is a consequence of the fact that, depending on the feasible collection $\XX'$, it may or may not be possible that alternative $a$ is selected with certainty from a menu $A$. More specifically, if $a$ appears in (at least) two feasible menus $A$ and $B$ then there is no $\theta$ for which $U_\theta(A) >U_\theta(B)$ and at the same time $\theta \Big( \big\{ z\in \RR^X :~ z_a>z_b~~ \forall b\in A\setminus\{a\} \big\}\Big)=1$. If $a$ is contained in only one menu, then clearly such $\theta$ does exist.

\section{Final comments}

This paper studies properties of various models of stochastic choice when available data is limited to the marginal distributions of menus and choices. We focused on three prominent models: The random utility model, the Luce model, and the independent random consideration set model. As can be expected, refuting these models based solely on the marginals is typically hard. But, somewhat surprisingly, one may be able to infer quite a lot about the models' parameters. For example, one can completely recover the value of alternatives for the Luce rule from marginal stochastic choice data. In addition, we studied the testable implications of endogenous menu choice in the temptation and self-control model of \cite{gul2001temptation} and the preference for flexibility model of \cite{kreps1979representation}. 

One can consider additional, perhaps more restrictive, models of stochastic choice and study their implications for the marginals. For instance, we have seen in Proposition \ref{prop-RUM} that the RUM places no restrictions on these distributions beyond those implied by unrestricted stochastic choice. But there are cases in which the set of alternatives has additional structure, and with this structure come natural restrictions on the possible preferences in the population. One example is when the alternatives are naturally ordered, and preferences are assumed to satisfy either single crossing or single peakedness \citep{apesteguia2017single}. A second example is when the alternatives are lotteries and preferences are assumed to be represented by expected utility \citep{gul2006random}. Third, in \cite{kitamura2018nonparametric} alternatives are bundles of goods and preferences satisfy monotonicity. A slightly different example is \cite{filiz2020progressive}, where agents may be `boundedly rational'. It would be interesting to see what are the implications of such domain restrictions for the marginals $(\mu,\lambda)$.

Another interesting direction would be to study two-period models in which an agent chooses a distribution over menus as well as a distribution over alternatives from each menu. These distributions may arise as the result of maximization behavior that trades off the benefits and costs of distributions of menus/alternatives similar to perturbed utility models such as \cite{fudenberg2015stochastic} and \cite{allen2022revealed}. If the maximization problems across the two periods are connected, then second-period stochastic choices would be constrained by the menu distribution chosen in the first period. That could generate non-trivial testable implications for the pair $(\mu,\lambda)$. 

Finally, while we assumed throughout that the observables are $\mu$ and $\lambda$, in applications it may be that more or less data can be accessed. One plausible scenario is that the researcher cannot observe how often each menu is available. Instead of observing $\mu$, the data only shows the availability of each alternative $a\in X$, namely, only $\xi(a)=\sum_{A:a\in A} \mu(A)$ for each $a$ is observable. Say that the marginal distribution of choices $\lambda\in \Delta(X)$ is potentially--rationalizable given $\xi\in [0,1]^X$ when there exists $\mu\in \Delta(\XX)$ such that $(\mu,\lambda)$ is rationalizable and such that $\xi(a)=\sum_{A:a\in A} \mu(A)$ for every $a\in X$. The following is a simple consequence of Proposition \ref{prop-basic}, see the Appendix for a proof.

\begin{corollary}\label{coro-no-correlation}
The distribution of choices $\lambda\in \Delta(X)$ is potentially--rationalizable given $\xi\in [0,1]^X$ if and only if $\lambda(a)\le \xi(a)$ for every $a\in X$.
\end{corollary}

It is also plausible that, in addition to the aggregate data, the researcher has choice frequencies in some menus (e.g., sales data from some retailers) but not in others. Or it may be that conditional choice probabilities for a particular alternative are available from its producer, while for the rest of the alternatives only the aggregate is known. It appears that much of our analysis would carry over to such situations, but we do not pursue these directions here and leave them to future work.

\bibliographystyle{plainnat}
\bibliography{Stochastic_choice}

\appendix

\section{Proofs}

\bigskip

\noindent \textbf{Proof of Proposition \ref{prop-RUM-inferior}}

\smallskip

That (1) implies (2) is obvious. We'll show first that (2) implies (3). Suppose that $\nu \in \Delta(O)$ RUM-rationalizes $(\lambda,\mu)$ and that $A$ is $\nu$-inferior. It follows that  
\begin{eqnarray*}
\sum_{a\in A} \lambda(a) &=& \sum_{a\in A} \sum_{\{B: a\in B\}}\mu(B) \nu(T[B,a]) = \sum_{a\in A} \sum_{\{B: a\in B\subseteq A\}}\mu(B) \nu(T[B,a]) =\\
& & \sum_{\{B: B\subseteq A\}}\mu(B) \sum_{a\in B} \nu(T[B,a]) = \sum_{\{B: B\subseteq A\}}\mu(B) =v_\mu(A),
\end{eqnarray*}
where the sfirst equality holds since $\nu$ RUM-rationalizes $(\mu,\lambda)$, the second equality follows from $\nu(T[B,a])=0$ whenever $a\in B\nsubseteq A$ since $A$ is $\nu$-inferior, the next is just a change in the order of summation, and the next follows from $\sum_{a\in B} \nu(T[B,a]) =1$ for every set $B$.

Finally, we show that (3) implies (1). Suppose $\lambda(A)=v_\mu(A)$ and that $\nu \in \Delta(O)$ RUM-rationalizes $(\mu,\lambda)$. Similar to the previous paragraph, we have that 
\begin{eqnarray*}
\sum_{a\in A} \lambda(a) = \sum_{a\in A} \sum_{\{B: a\in B\}}\mu(B) \nu(T[B,a]) \ge \sum_{a\in A} \sum_{\{B: a\in B\subseteq A\}}\mu(B) \nu(T[B,a]) =v_\mu(A).
\end{eqnarray*}
By assumption, the first and last expressions are equal, so the inequality must in fact hold as equality. This implies that $\mu(B) \nu(T[B,a])=0$ for every pair $a,B$ with $a\in A\cap B$ and $B\nsubseteq A$. In particular, for any pair of alternatives $a\in A$ and $b\notin A$ we have that $\mu(\{a,b\})\nu(T[\{a,b\},a])=0$. By assumption $\mu(\{a,b\})>0$, which implies that any $\succ$ with $a\succ b$ is not in the support of $\nu$, proving (1).

\bigskip

\noindent \textbf{Proof of Proposition \ref{prop-homeomorphism}}

\smallskip


We start by showing that the mapping $u \longrightarrow \lambda$ given by (\ref{eqn-Luce}) is one-to-one. Suppose that $u,u'$ induce the same marginal distribution of choices $\lambda$. Let $a^*\in\argmax_{a \in X} \frac{u(a)}{u'(a)}$. For every menu $A$ that contains $a^*$ we have
$$\frac{u(a^*)}{\sum_{a \in A} u(a)} = \frac{1}{\sum_{a\in A} \frac{u(a)}{u(a^*)}} \ge \frac{1}{\sum_{a\in A} \frac{u'(a)}{u'(a^*)}} = \frac{u'(a^*)}{\sum_{a \in A} u'(a)},$$
where the inequality follows since $\frac{u(a)}{u(a^*)} \le \frac{u'(a)}{u'(a^*)}$ for every $a\in X$. Since by assumption 
$$\sum_{\{A:~ a^*\in A\}} \mu(A) \frac{u(a^*)}{\sum_{a \in A} u(a)} = \lambda(a^*) = \sum_{\{A:~ a^*\in A\}} \mu(A) \frac{u'(a^*)}{\sum_{a \in A} u'(a)},$$
it follows that for every menu $A$ in the support of $\mu$ we have$\frac{u(a^*)}{\sum_{a \in A} u(a)} = \frac{u'(a^*)}{\sum_{a \in A} u'(a)}$. This implies in turn that $\frac{u(a)}{u(a^*)} = \frac{u'(a)}{u'(a^*)}$ for all $a\in A$ whenever $A$ is in the support of $\mu$. By assumption, for every $a\neq a^*$, there is $A$ in the support of $\mu$ such that $\{a,a^*\}\subseteq A$. Thus, the ratio $\frac{u(a)}{u'(a)}$ is constant in $a$. Since both $u$ and $u'$ are in $\Delta(X)$, it follows that $u=u'$. 

Next, we argue that under the assumption of the proposition $Core(v_\mu)$ is a set of full dimension in $\Delta(X)$ and that if $(\mu,\lambda)$ is Luce rationalizable then $\lambda\in Int(Core(v_\mu))$. Indeed, fix some $u\in Int(\Delta(X))$ and let $\lambda$ be its image. Recall that $\sum_{a \in A} \lambda(a)=v_\mu(A)$ holds if and only if every element of $A$ is never chosen when some element of $A^c$ is available. Given menu $A\neq X$, consider some $a\in A$ and $b\in A^c$. By assumption there is $B\in \XX$ with $\mu(B)>0$ and $\{a,b\}\subseteq B$. When $B$ is the available menu, alternative $a$ is selected with positive probability of $\frac{u(a)}{\sum_{c \in B}u(c)}$. Thus, $\sum_{a \in A} \lambda(a)>v_\mu(A)$ for every menu $A$, as needed.

The rest of the proof follows the footsteps of the proof of Lemma 4 in \cite{monderer1992weighted}. The mapping $u \longrightarrow \lambda$ given by (\ref{eqn-Luce}) maps $Int(\Delta(X))$ into $Int(Core(v_\mu))$ and is clearly continuous. We have argued that this map is injective and that the domain and range have the same dimension. By the Invariance of Domain theorem the image of $Int(\Delta(X))$ is open in $Int(Core(v_\mu))$. 

To finish the proof, consider the set of $\lambda$'s in $Int(Core(v_\mu))$ that are \emph{not} in the image. We claim that this set is also open in $Int(Core(v_\mu))$. Indeed, if it is not then there is $\lambda_0$ in this set and a sequence $\lambda_n \to \lambda_0$ such that each $\lambda_n$ is the image of some $u_n$ under Equation (\ref{eqn-Luce}). By compactness we may assume w.l.o.g. that $u_n$ converges in $\Delta(X)$, say to $u_0$. By continuity, $\lambda_0$ is the image of $u_0$, so it must be that $u_0$ is on the boundary of $\Delta(X)$. This implies that $\lambda_0(a)=0$ for some alternative $a$, contradicting the assumption that $\lambda_0\in Int(Core(v_\mu))$. Therefore, since $Int(Core(v_\mu))$ is connected, it must be that the image of $Int(\Delta(X))$ under Equation (\ref{eqn-Luce}) is the entire set $Int(Core(v_\mu))$.

\bigskip

\noindent \textbf{Proof of Proposition \ref{prop-Luce-symmetry}}

\smallskip

Given $a,b$, define the collections of menus $F(a,b) = \{A\in \XX ~:~ a\in A,~ b\notin A\}$ and $F(b,a) = \{A\in \XX ~:~ b\in A,~ a\notin A\}$. Let $g:F(a,b)\to F(b,a)$ be defined by $g(A)=(A\cup\{b\})\setminus\{a\}$. Then clearly $g$ is a bijection, and $\mu(A)=\mu(g(A))$ for every $A\in F(a,b)$ by $ab$-exchangeability. 

Now, suppose that $u$ Luce rationalizes the marginal stochastic choice dataset $(\mu,\lambda)$. It follows that,  
\begin{equation}\label{eqn-both-sums}
\lambda(a)= \sum_{\{A: a\in A\}} \mu(A) \frac{u(a)}{\sum_{c \in A} u(c)} = \sum_{\{A: \{a,b\}\subseteq A\}} \mu(A) \frac{u(a)}{\sum_{c \in A} u(c)} + \sum_{A\in F(a,b)} \mu(A) \frac{u(a)}{\sum_{c \in A} u(c)}.
\end{equation}
If $u(a)\ge u(b)$, then the first sum satisfies  
\begin{equation}\label{eqn-first-sum}
    \sum_{\{A: \{a,b\}\subseteq A\}} \mu(A) \frac{u(a)}{\sum_{c \in A} u(c)} \ge \sum_{\{A: \{a,b\}\subseteq A\}} \mu(A) \frac{u(b)}{\sum_{c \in A} u(c)}.
\end{equation}
As for the second sum, we have 
\begin{eqnarray}\label{eqn-second-sum}
    \sum_{A\in F(a,b)} \mu(A) \frac{u(a)}{\sum_{c \in A} u(c)} & \ge & \sum_{A\in F(a,b)} \mu(A) \frac{u(b)}{\sum_{c \in g(A)} u(c)} = \nonumber \\ 
    & & \sum_{A\in F(a,b)} \mu(g(A)) \frac{u(b)}{\sum_{c \in g(A)} u(c)} = 
    \sum_{A\in F(b,a)} \mu(A) \frac{u(b)}{\sum_{c \in A} u(c)},
\end{eqnarray}
where the inequality is since $t\to \frac{t}{t+c}$ is increasing (when $c>0$), the first equality follows from $\mu(A)=\mu(g(A))$ for all $A\in F(a,b)$, and the last equality follows from $g$ being a bijection. Combining (\ref{eqn-both-sums}), (\ref{eqn-first-sum}), and (\ref{eqn-second-sum}) we get 
$$\lambda(a)\ge \sum_{\{A: \{a,b\}\subseteq A\}} \mu(A) \frac{u(b)}{\sum_{c\in A} u(c)} + \sum_{A\in F(b,a)} \mu(A) \frac{u(b)}{\sum_{c\in A} u(c)} = \lambda(b).$$
The converse, that $\lambda(a)\ge \lambda(b)$ implies $u(a)\ge u(b)$, easily follows.

\bigskip

\noindent \textbf{Proof of Proposition \ref{prop-RCS}}

\smallskip

Suppose that $(\mu,\lambda)$ is $\succ$-IRCS rationalizable, and let $\gamma=(\gamma_a)$ satisfy Equation (\ref{eqn-RCS}). We can rewrite Equation (\ref{eqn-RCS}) as
$$\gamma_a = \lambda(a) \left(\sum_{\{A^*\in \XX^*:~ a\in A^*\}} \mu(A^*) \left[\prod_{\{b\in A:~ b\succ a\}}(1-\gamma_b)\right]\right)^{-1},$$
where this is well-defined due to the assumption on $\mu$. For $a_1$ this gives 
$$\gamma_{a_1} = \lambda(a_1)\left[\mu(\{A^*\in\XX^*:~ a_1\in A^*\})\right]^{-1}=t^\succ_1.$$ 
For $a_2$ we have
$$\gamma_{a_2} = \lambda(a_2)\Big[\mu(\{A^*\in\XX^*:~ A^*\cap \{a_1,a_2\}=\{a_1,a_2\}\})(1-\gamma_{a_1}) + \mu(\{A^*\in\XX^*:~ A^*\cap \{a_1,a_2\}=\{a_2\}\})\Big]^{-1}.$$
Since we already established that $\gamma_{a_1} =t^\succ_1$, we get that the latter is also equal to 
$$\lambda(a_2)\left[\mu(\{A^*\in\XX^*:~ A^*\cap \{a_1,a_2\}=\{a_1,a_2\}\})(1-t^\succ_1) + \mu(\{A^*\in\XX^*:~ A^*\cap \{a_1,a_2\}=\{a_2\}\})\right]^{-1},$$
which is precisely the definition of $t^\succ_2$. Thus, $\gamma_{a_2}=t^\succ_2$. Continuing by induction it easily follows that $\gamma_{a_k}=t^\succ_k$ for all $k$, so that $t^\succ_k \le 1$ is satisfied. This also shows that the rationalizing probabilities $\gamma$ are uniquely determined by $(\mu,\lambda)$ and $\succ$.

The proof of the converse is similar. Setting $\gamma_{a_k}=t^\succ_k$ for every $k$ shows that $(\mu,\lambda)$ is $\succ$-IRCS rationalizable when $t^\succ_k \le 1$. We omit the details.

\bigskip

\noindent \textbf{Proof of Corollary \ref{coro-RCS2}}

\smallskip

Given a full support $\mu$ and $\succ\in O$, let $\Lambda_{(\succ,\mu)}$ be the set of all marginal choice distributions $\lambda$ such that $(\mu,\lambda)$ is $\succ$-IRCS rationalizable. Let $\lambda_\succ\in \Lambda_{(\succ,\mu)}$ be the choice distribution induced by $\succ$ and by $\gamma\equiv 1$ from Equation (\ref{eqn-RCS}). We first argue that $\lambda_\succ \notin \Lambda_{(\succ',\mu)}$ for every $\succ'\neq \succ$. Indeed, the full support assumption immediately implies that to rationalize $\lambda_\succ$ it is necessary that $\gamma \equiv 1$. Furthermore, Lemma \ref{lemma-convex} and Proposition \ref{prop-core-convex} imply that $\lambda_\succ \neq \lambda_{\succ'}$ whenever $\succ\neq\succ'$. Thus, $(\succ,\gamma\equiv 1)$ is the unique pair that IRCS-rationalizes $(\mu,\lambda_\succ)$.

Next, we argue that if $\{\lambda^m\}_m$ is a sequence of choice distributions such that $\lambda^m(x^*)\to 0$ and such that $\lambda^m\in\Lambda_{\succ,\mu}$ for every $m$, then $\lambda^m\to \lambda_\succ$. In other words, if we have a sequence where the probability of the default option converges to zero, then we converge to a full consideration model where $\gamma \equiv 1$. Indeed, the full support assumption implies that if $(\succ, \gamma^m)$ rationalizes $\lambda^m$, then we must have $\gamma^m_a\to 1$ for every $a$ in order for $\lambda^m(a^*)$ to vanish. But $\gamma^m_a\to 1$ for every $a$ implies that $\lambda^m\to \lambda_\succ$ by continuity of Equation (\ref{eqn-RCS}).

Now, Proposition~\ref{prop-RCS} implies that each $\Lambda_{(\succ',\mu)}$ is closed from the weak inequalities $t^{\succ'}_k \le 1$, and the first paragraph above implies that $\lambda_\succ \notin \bigcup_{\succ'\neq \succ} \Lambda_{(\succ',\mu)}$. Let $\epsilon_\succ>0$ be small enough so that the ball of radius $\epsilon_\succ$ around $\lambda_\succ$ doesn't intersect the above union. Then by the second paragraph of this proof there is $\delta_\succ>0$ such that if $\lambda\in \Lambda_{(\succ,\mu)}$ and $\lambda(a^*)<\delta_\succ$ then $\|\lambda-\lambda_\succ\|<\epsilon_\succ$. Setting $\bar\delta=\min_\succ \delta_\succ$ completes the proof.

\bigskip

\noindent \textbf{Proof of Lemma \ref{lemma-TSC-1}}

\smallskip

Suppose first that $a\in \bar A = A \setminus \bigcup_{\{B \in \XX': B \subsetneq A \} } B$. Define $u,v$ by $u(a)=2$, $u(b)=0$ for all $b\neq a$, $v(b)=0$ for all $b\in A$, and $v(b)=1$ for all $b\notin A$. First, we have that $u(a)+v(a)=2$ while $u(b)+v(b)\le 1$ for any other $b$, so that in particular $u(a)+v(a)>u(b)+v(b)$ for all $b\in A\setminus\{a\}$. Second, 
$$U_{u,v}(A)=\max_{a'\in A}[u(a')+v(a')]-\max_{b\in A}[v(b)] = 2-0=2.$$ 
Let $B\in \XX'$ be any other feasible menu and consider two possible cases: If $a\notin B$ then 
$$U_{u,v}(B)= \max_{a'\in B}[u(a')+v(a')]-\max_{b\in B}[v(b)] \le 1-0=1;$$
if $a\in B$ then by assumption $B$ must contain an element from $A^c$, so that 
$$U_{u,v}(B)= \max_{a'\in B}[u(a')+v(a')]-\max_{b\in B}[v(b)] \le 2-1=1.$$
It follows that in either case $U_{u,v}(A)>U_{u,v}(B)$, as needed.

Conversely, suppose that $a\in B\subsetneq A$ for some $B\in \XX'$. Consider any pair of functions $(u,v)$ that satisfies $u(a)+v(a)>u(b)+v(b)$ for all $b\in A\setminus\{a\}$. Then in particular $u(a)+v(a)>u(b)+v(b)$ for all $b\in B\setminus\{a\}$, so that
$$U_{u,v}(B)= \max_{a'\in B}[u(a')+v(a')]-\max_{b\in B}[v(b)] = [u(a)+v(a)] - \max_{b\in B}[v(b)] \ge [u(a)+v(a)] - \max_{b\in A}[v(b)] = U_{u,v}(A).$$
Therefore, it can't be that $U_{u,v}(A)>U_{u,v}(B)$ for all $B\in \XX'\setminus\{A\}$.

Finally, we need to show that if $A\in R(\XX')$ then there is no pair $(u,v)$ such that $U_{u,v}(A)>U_{u,v}(B)$ for any $B\in \XX'\setminus\{A\}$. Fix $A\in R(\XX')$ and a pair $(u,v)$. Let $a^*\in \argmax_{a\in A} u(a)+v(a)$. By assumption, there is $B\in \XX'$ such that $a^*\in B\subsetneq A$. Then as in the previous paragraph we have that 
$$U_{u,v}(B) = [u(a^*)+v(a^*)] - \max_{b\in B}[v(b)] \ge [u(a^*)+v(a^*)] - \max_{b\in A}[v(b)] = U_{u,v}(A),$$
where the first equality is because $a^*\in \argmax_{a\in B} u(a)+v(a)$, and the inequality is due to $B\subseteq A$.
\bigskip

\noindent \textbf{Proof of Lemma \ref{lemma-TSC-2}}

\smallskip

Suppose first that $\pi$ satisfies the requirements of the lemma. By Lemma \ref{lemma-TSC-1}, for every $A\in NR$ and every $a\in \bar A$ there is a pair $u,v$ such that $U_{u,v}(A)>U_{u,v}(B)$ for all $B\in \XX'\setminus\{A\}$ and such that $u(a)+v(a)>u(b)+v(b)$ for all $b\in A\setminus\{a\}$. Denote this pair by $(u,v)^{A,a}$. Let $\psi$ be the distribution with support $\left\{(u,v)^{A,a}\right\}_{A\in NR,~ a\in \bar A}$, and with
$\psi\left((u,v)^{A,a}\right) = \mu(A)\pi(a|A)$ for every $a,A$. Then it is immediate that $\psi$ satisfies conditions (1) and (2) of Definition \ref{def-gul-pese} and therefore that $(\mu,\lambda)$ is $\XX'$-TSC rationalizable.

Conversely, suppose that $(\mu,\lambda)$ is $\XX'$-TSC rationalizable and let $\psi$ be a rationalizing distribution over pairs $u,v$. First, Lemma \ref{lemma-TSC-1} implies that if $A\in R(\XX')$ then $\mu(A)=0$. Second, for any $A\in NR$ 
define $\pi(\cdot|A)$ as follows. If $\mu(A)=0$, then $\pi(\cdot|A)$ is any distribution with support contained in $\bar A$; and if $\mu(A)>0$ then 
\begin{eqnarray*}
\pi(a|A) = \frac{1}{\mu(A)} \psi \Big(\big\{ (u,v) & : & u(a)+v(a)>u(b)+v(b)~~ \forall b\in A\setminus\{a\} \textit{ and }\\
& & U_{u,v}(A)>U_{u,v}(B)~~ \forall B\in \XX'\setminus \{A\}\big\}\Big).
\end{eqnarray*}
Then clearly $\pi$ satisfies the equality in the lemma and it follows from Lemma \ref{lemma-TSC-1} that $\pi(a|A)=0$ whenever $a\notin \bar A$.



\bigskip

\noindent \textbf{Proof of Proposition \ref{prop-flow-cut}}

\smallskip

\noindent `Only if': Suppose that there is $\pi=\{\pi(\cdot | y)\in \Delta(h(y))\}_y$ such that $\lambda(a) = \sum_y \mu(y)\pi(a|y)$ for every $a\in X$ and fix a menu $A\in \XX$. Then
\begin{eqnarray*}
\sum_{a\in A} \lambda(a) =  \sum_{a\in A} \sum_{y} \mu(y)\pi(a|y) \ge  \sum_{\{y:~ h(y)\subseteq A\}} \mu(y) \sum_{a\in A} \pi(a|y)  = \sum_{\{y:~ h(y)\subseteq A\}} \mu(y),
\end{eqnarray*}
so that the required collection of inequalities is satisfied.

\smallskip

\noindent `If': Consider a directed bipartite graph with sets of nodes $V_1=X$ and $V_2=Y$, and where there is an edge from $a\in V_1$ to $y\in V_2$ if and only if $a\in h(y)$. Add two additional nodes, a source and a sink, denoted $s$ and $t$ respectively. Also add edges from $s$ to every $a\in V_1$ and from every $y\in V_2$ to $t$. Denote by $G=(V,E)$ the resulting augmented graph. For every $e\in E$ set a capacity $c(e)\in \RR_+$ as follows: If $a\in X$ then $c(s,a)=\lambda(a)$; If $y\in Y$ then $c(y,t)=\mu(y)$; and $c(a,y)=1$ for every edge in the original bipartite graph.

We claim that, under the assumption of the proposition, any cut $E'\subseteq E$ that separates $s$ from $t$ has total capacity of at least 1. This is clearly the case if one of the $(a,y)$ edges is in $E'$; if not, then for every $a\in X$ either $(s,a)\in E'$ or $\{(y,t)~:~ a\in h(y)\}\subseteq E'$ (or both). Let $A_0 = \{a\in X ~:~ (s,a)\in E'\}$. Then the total capacity of $E'$ satisfies
\begin{align*}
\sum_{e\in E'} c(e) &\ge \sum_{a\in A_0} c(s,a) + \sum_{\{y :~ h(y)\nsubseteq A_0\}} c(y,t) \\
&= \sum_{a \in A_0} \lambda(a) + 1- \sum_{\{y :~ h(y)\subseteq A_0\}} c(y,t) \\
&= \sum_{a \in A_0} \lambda(a)+1-\sum_{\{y:~ h(y)\subseteq A_0\}}\mu(y)\\
&\ge 1,
\end{align*}
where the final inequality is by the assumption of the proposition.

Clearly, there exists a cut $E'$ of capacity equal to 1. Thus, the optimal value of the min-cut program for $G$ is 1. By the max-flow min-cut theorem of linear programming (see, for example, Theorem 7.13 in \cite{kleinberg2006algorithm}), there exists a flow $f:E\to\RR_+$ such that $\sum_{a\in X} f(s,a)=\sum_{y\in Y} f(y,t)=1$. In particular, $f(s,a)=\lambda(a)$ and $f(y,t)=\mu(y)$ for all $a$ and $y$. For every edge $(a,y)$ define $\pi(a|y)=\frac{f(a,y)}{\mu(y)}$ (and $\pi(\cdot |y)$ arbitrarily on $h(y)$ when $\mu(y)=0$). For every $a\in X$ we thus have
$$\lambda(a) = f(s,a) = \sum_y f(a,y) = \sum_y \mu(y)\pi(a|y),$$
where the second equality is by the flow conservation constraints at the node $a$, and the last equality is by the definition of $\pi(a|y)$. This completes the proof.

\bigskip

\noindent \textbf{Proof of Proposition \ref{prop-kreps}}

\smallskip

Suppose first that $(\mu,\lambda)$ is $\XX'$-PF rationalizable and let $\psi$ be the rationalizing distribution. If $A,B\in \XX'$ and $A\subsetneq B$ then for any $\theta$ we have 
$$U_\theta(A)=\EE\Big(\max_{a\in A} Z_a\Big) \le \EE\Big(\max_{a\in B} Z_a\Big) = U_\theta(B),$$
implying that $\Theta_{\XX'}(A)=\emptyset$ and hence that $\mu(A)=\psi(\Theta_{\XX'}(A))=0$. Therefore, $\mu(\tilde R(\XX'))=0$. In addition, it follows from Proposition \ref{prop-basic} that $\lambda\in Core(v_\mu)$.

\medskip

Conversely, fix some $A\notin \tilde R(\XX')$, an alternative $a\in A$, and a positive integer $n$. Consider the distribution $\theta^{A,a}_n$ defined as follows: With probability $\frac{n-1}{n}$ the realized vector $z$ is such that $z_{a}=1$ and $z_b=0$ for any other alternative $b\in X$; and with probability $\frac{1}{n}$ one of the other alternatives $a'\in A$ is randomly and uniformly selected and the realized vector $z$ is such that $z_{a'}=1$ and $z_b=0$ for all other alternatives $b\in X$. We have that $U_\theta(A)=1$, and since $A\notin \tilde R(\XX')$ we also have $U_\theta(B)<1$ for every feasible $B\neq A$. Thus, if $\psi$ is the Dirac measure on $\theta^{A,a}_n$ then the resulting pair $(\mu,\lambda)$ is given by $\mu(A)=1$, $\lambda(a)=\frac{n-1}{n}$, and $\lambda(a')=\frac{1}{n(|A|-1)}$ for every $a'\in A\setminus\{a\}$. 

Next, fix some $\mu$ with $\mu(\tilde R(\XX'))=0$ and denote by $C(\mu)$ the set of $\lambda$'s such that $(\mu,\lambda)$ is $\XX'$-PF rationalizable. Note that $C(\mu)$ is convex (by the linearity of $\lambda$ in $\psi$). Fix an ordering $\succ\in O$ and let $\lambda_\succ$ be the corresponding extreme point of $Core(v_\mu)$ associated with $\succ$. Let $\psi_n$ be the distribution that assigns probability $\mu(A)$ to $\theta^{A,a}_n$, where $a$ is the top element of $A$ according to $\succ$, and denote $(\mu_n,\lambda_n)$ the pair induced by $\psi_n$. Then $\mu_n=\mu$ for all $n$, and $\lambda_n\longrightarrow \lambda_\succ$. Therefore, every extreme point of $Core(v_\mu)$ is in the closure $C(\mu)$. It follows that the closure of $C(\mu)$ is equal to $Core(v_\mu)$. By \citet[Corollary 6.3.1]{rockafellar1970convex} the relative interior of $Core(v_\mu)$ is contained in $C(\mu)$, and the proof is complete.

\bigskip

\noindent \textbf{Proof of Corollary \ref{coro-no-correlation}}

\smallskip

First, it is obvious that if $\lambda$ is potentially--rationalizable given $\xi$ then $\lambda(a)\le \xi(a)$ for every $a\in X$. For the converse, fix $\xi$ and $\lambda$ that satisfy these inequalities. We describe an algorithm that stops after a finite number of steps and produces the desired $\mu$.

Label the alternatives arbitrarily, $X=\{a_1,\ldots,a_n\}$. Given any $\mu\in \Delta(\XX)$, let $\xi_\mu\in [0,1]^X$ be defined by $\xi_\mu(a_i)=\sum_{\{A:~ a_i\in A\}} \mu(A)$, and let $D_\mu=\{1\le i \le n:~ \xi_\mu(a_i)< \xi(a_i)\}$ be the set of alternatives for which the probability of availability based on $\mu$, $\xi_\mu(a_i)$, is strictly below the required level $\xi(a_i)$. 

The algorithm starts with $\mu^0$ whose support is just the singletons $\{a_i\}_{i=1}^n$, and where $\mu^0(a_i)=\lambda(a_i)$ for every $i$. Note that, by assumption, $\xi_{\mu^0}\le \xi$. 

For some $k=1,2,\ldots$ suppose that a distribution $\mu^{k-1}$ is given. If $D_{\mu^{k-1}}=\emptyset$ then stop. Otherwise, we now describe how to obtain $\mu^k$. 

Let $i^* = \min D_{\mu^{k-1}}$ so that in particular $\xi_{\mu^{k-1}}(a_{i^*}) <  \xi(a_{i^*})$. We claim that there must exist a menu $\tilde A$ in the support of $\mu^{k-1}$ such that $a_{i^*}\notin \tilde A$. Indeed, if $a_{i^*}$ is in every menu in the support of $\mu^{k-1}$ then $\xi_{\mu^{k-1}}(a_{i^*})=1$, contradicting $\xi_{\mu^{k-1}}(a_{i^*}) <  \xi(a_{i^*})$. Let $t=\min\{\mu(\tilde A), \xi(a_{i^*}) - \xi_{\mu^{k-1}}(a_{i^*})\}>0$. Define $\mu^k$ by $\mu^k(\tilde A) = \mu^{k-1}(\tilde A)-t$, $\mu^k(\tilde A\cup\{a_{i^*}\}) = \mu^{k-1}(\tilde A\cup\{a_{i^*}\})+t$, and $\mu^k(A) = \mu^{k-1}(A)$ for every other menu $A$. In words, a mass of $t$ is moved from the menu $\tilde A$ to the menu $\tilde A\cup\{a_{i^*}\}$.

We have that $\xi_{\mu^k}(a_{i^*}) = \xi_{\mu^{k-1}}(a_{i^*})+t$, and $\xi_{\mu^k}(a_i) = \xi_{\mu^{k-1}}(a_i)$ for every $i\neq i^*$. Therefore, $\xi_{\mu^k}$ is an increasing sequence, bounded above by $\xi$. Moreover, if $i^*$ is not removed from $D_{\mu^{k}}$, then $\tilde A$ is removed from the support of $\mu^k$, implying that after at most $2^{n-1}$ iterations an equality $\xi_{\mu^k}(a_{i^*}) = \xi(a_{i^*})$ is achieved. It follows that the algorithm terminates after at most $n\times 2^{n-1}$ iterations.

Let $\mu^*$ be the resulting final distribution. By construction, $\xi_{\mu^*}(a_i) = \xi(a_i)$ for every $i$. It is left to show that $(\mu^*,\lambda)$ is rationalizable. Note however that $\lambda\in Core(v_{\mu^0})$ and that $v_{\mu^k}\le v_{\mu^{k-1}}$ for every $k$ since $\mu^k$ is obtained from $\mu^{k-1}$ by shifting mass from a set to one of its supersets. Thus, $\lambda\in Core(v_{\mu^0}) \subseteq Core(v_{\mu^*})$, so we are done by Proposition \ref{prop-basic}.

\end{document}